\documentclass[review]{elsarticle}

\usepackage{lineno,hyperref}

\journal{\ac{NIM}}

\usepackage[english]{babel}
\usepackage{textcomp,gensymb}
\usepackage[utf8x]{inputenc}
\usepackage[]{units}
\usepackage{acronym}
\usepackage[T1]{fontenc}
\usepackage{xcolor}
\usepackage{lettrine} 
\usepackage{ upgreek }
\usepackage{gensymb} 
\usepackage{float}
\usepackage{placeins}
\usepackage{nomencl}
\usepackage{caption}
\usepackage{subcaption}
\usepackage{ragged2e}
\usepackage{multirow}
\makeatletter
\newcommand{\ssymbol}[1]{^{\@fnsymbol{#1}}}
\makeatother
\usepackage{amsmath}
\usepackage{graphicx}
\usepackage[colorinlistoftodos]{todonotes}
\usepackage{hyperref} 
\usepackage{booktabs,caption}
\usepackage[flushleft]{threeparttable}
\hypersetup{colorlinks = true,linkcolor = blue,anchorcolor =red,citecolor = blue,filecolor = red,urlcolor = red}
\graphicspath{ {./img/} }
\acrodef{LEMMA}[LEMMA]{Low EMittance Muon Accelerator}
\acrodef{MAP}[MAP]{Muon Accelerator Program}

\begin{document}

\begin{frontmatter}

\title{Muon detection in electron-positron annihilation  for muon collider studies\tnoteref{mytitlenote}}


\author[Torino,INFN_TO]{N. Amapane}
\author[INFN_LNF]{M. Antonelli}
\author[INFN_RM]{F. Anulli}
\author[Insubria,INFN_MIB]{G. Ballerini}
\author[INFN_FE]{L. Bandiera}
\author[INFN_TO]{N. Bartosik}
\author[INFN_RM]{M. Bauce}
\author[INFN_PD]{A. Bertolin}
\author[INFN_TO]{C. Biino}
\author[INFN_LNF]{O. R. Blanco-Garc\'ia}
\author[INFN_LNF]{M. Boscolo}
\author[Insubria,INFN_MIB]{C. Brizzolari}
\author[Torino,INFN_TO]{A. Cappati}
\author[Sapienza,INFN_RM]{F. Casaburo \corref{mycorrespondingauthor}}
\ead{fausto.casaburo@uniroma1.it}
\cortext[mycorrespondingauthor]{Corresponding author}
\author[INFN_TS]{M. Casarsa}
\author[Sapienza,INFN_RM]{G. Cavoto}
\author[Sapienza,INFN_RM]{G. Cesarini}
\author[INFN_RM]{F. Collamati}
\author[Torino,INFN_TO]{G. Cotto}
\author[INFN_PD]{C. Curatolo}
\author[CERN]{R. Di Nardo}
\author[INFN_PD]{F. Gonella}
\author[Padova,INFN_PD]{S. Hoh}
\author[INFN_LNF]{M. Iafrati}
\author[INFN_RM]{F. Iacoangeli}
\author[INFN_TO]{B. Kiani}
\author[Sapienza]{R. Li Voti}
\author[CERN,INFN_PD]{D. Lucchesi}
\author[Insubria,INFN_MIB]{V. Mascagna}
\author[CERN,INFN_PD]{A. Paccagnella}
\author[INFN_TO]{N. Pastrone}
\author[CERN,INFN_PD]{J. Pazzini}
\author[Torino,INFN_TO]{M. Pelliccioni}
\author[INFN_LNF]{B. Ponzio}
\author[Insubria,INFN_MIB]{M. Prest}
\author[INFN_LNF]{M. Ricci}
\author[INFN_RM]{S. Rosati}
\author[Padova,INFN_PD]{R. Rossin}
\author[INFN_LNF]{M. Rotondo}
\author[Torino,INFN_TO]{O. Sans Planell}
\author[INFN_PD]{L. Sestini}
\author[Insubria,INFN_MIB]{M. Soldani}
\author[Curien]{A. Triossi}
\author[INFN_MIB]{E. Vallazza}
\author[INFN_PD]{S. Ventura}
\author[Padova,INFN_PD]{M. Zanetti}

\address[Torino]{Università degli Studi di Torino- Torino, Italy}
\address[INFN_TO]{INFN Sezione Torino- Torino, Italy}
\address[INFN_LNF]{INFN Laboratori Nazionali di Frascati - Frascati, Italy}
\address[INFN_RM]{INFN Sezione Roma- Rome, Italy}
\address[Insubria]{Università degli Studi dell’Insubria - Como, Italy}
\address[INFN_MIB]{INFN Sezione di Milano Bicocca - Milan, Italy}
\address[INFN_FE]{INFN Sezione di Ferrara - Ferrara, Italy}
\address[INFN_PD]{INFN Sezione di Padova - Padova, Italy}
\address[Sapienza]{Sapienza Università di Roma- Roma, Italy}
\address[INFN_TS]{INFN Sezione di Trieste - Trieste, Italy}
\address[CERN]{CERN - Geneva, Switzerland}
\address[Padova]{Università di Padova - Padova, Italy}
\address[Curien]{Institut Pluridisciplinaire Hubert Curien, Strasbourg, France}

\begin{abstract}
 The investigation of the energy frontier in physics requires novel concepts for future colliders. The idea of a muon collider is very appealing since it would allow to study particle collisions at up to tens of $\unit{TeV}$ energy, while offering a  cleaner experimental environment with respect to hadronic colliders. One key element in the muon collider design is the low-emittance muon production. Recently, the \ac{LEMMA} collaboration  has explored  the  muon pair production close to its kinematic threshold by annihilating $\unit[45]{GeV}$ positrons with electrons in a low $Z$ material target. In this configuration, muons are emerging from the target with a naturally low-emittance.  In this paper we describe the performance of a system, to study this production mechanism, that consists in several segmented absorbers with alternating active layers  {\color{black}{ composed of} } fast Cherenkov detectors together with a muon identification technique based on this detector.
Passive layers were made of tungsten.
We collected data corresponding to muon and electron beams produced at the H2 line in the North Area of the \ac{CERN} in September 2018.

\end{abstract}

\begin{keyword}
Muon Collider, Cherenkov detectors
\end{keyword}

\end{frontmatter}

\modulolinenumbers[5]
\tableofcontents

\section*{Introduction}
\addcontentsline{toc}{section}{Introduction}

Exploring the high energy frontier represents a great opportunity to investigate the fundamental laws of nature. This requires a particle collider able to accelerate elementary particles to the highest possible energy. A muon collider represents an appealing though challenging solution {\color{black}{for the aforementioned purpose}} \cite{Franceschini:2021oog, MAPsite}. Previous studies claimed that a muon collider is conceivable to reach the multi–$\unit{TeV}$ energy frontier with the possibility to study Higgs boson properties~\cite{MAPsite}. {\color{black}{Muons in a circular storage ring}} are in fact emitting much less {\color{black}{synchrotron radiation}} than electrons with the same energy; {\color{black}{ therefore, muons accelerated in a circular collider can reach a \ac{CM} energy higher than electrons in the same ring. However, the muon lifetime of approximately $\unit[2]{\mu s}$ is a limiting factor on the design of a muon accelerator complex}}, requiring a fast accelerating complex. {\color{black}{Moreover, reaching a small emittance }}is  one of the crucial aspects to achieve a high luminosity in a muon collider \cite{Long_muon_collider}. 

The \ac{LEMMA} project \cite{Antonelli:2016oog} aims to study the possibility of producing muons from the $e^{+}e^{-}$ annihilation process. A high intensity positron beam, with energy just above the $\unit[43.7]{GeV}$ threshold for muon pair production through the $e^{+}e^{-}\rightarrow\mu^{+}\mu^{-}$ process, impinging on a low $Z$ fixed target \cite{Cesarini2021} could produce muons with naturally small divergence, resulting in a low transverse emittance. This would avoid   the need of a subsequent beam cooling stage, otherwise required in a muon production scheme based on pion decays as considered in the \ac{MAP} \cite{Palmer:1996fzp}. Experimental data in the energy regime close to muon pair production threshold are not frequent since most of the measurements of the $e^{+}e^{-}\rightarrow\mu^{+}\mu^{-}$ process are performed at higher \ac{CM} energy values~\cite{Thomson}. It is therefore necessary to measure the aforementioned production cross section and the produced muon pair  kinematic properties for several values of the \ac{CM} energy near the muon pair production threshold to probe the validity of the predictions.

While the leading-order \ac{QED} $e^{+}e^{-}\rightarrow\mu^{+}\mu^{-}$ cross section calculation is well established, higher order radiative effects, due to Coulomb interaction, might gain importance close to the kinematic threshold when evaluating both the muon pair production rate and their angular distribution \cite{Bystritskiy:2005oog}. A dedicated experimental setup has been arranged to study the muon pair production through $\unit[45]{GeV}$ positrons impinging on a Beryllium or Carbon target in different data-taking campaigns~\cite{Amapane:2019oog}.

The aim of this paper is to describe a system made of two segmented and instrumented absorbers, that has been used to effectively study the $e^{+}e^{-}\rightarrow\mu^{+}\mu^{-}$ process near the muon pair production threshold region. 

These absorbers were part of the apparatus installed at the \ac{CERN} North Area beam lines during the 2017 and 2018 data-taking periods. These were deployed to identify both positive and negative muons and discriminate them from electrons and positrons. The absorbers were initially conceived as massive but portable devices that could be used as sampling calorimeters to study the development of hadronic showers initiated by the \ac{LHC} beam interactions \cite{Iacoangeli:2018fzp}. Their performances for muon identification have been tested using both electron and muon beams. Results regarding the performances of one of these absorbers, named \ac{HORSA} in the following, are described in Section \ref{sec:performance}. To validate these performances, we measured the ratio of the number of muons to the number of electrons in events produced from a $\unit[45]{GeV}$ positron beam data hitting a Carbon target and compared with a \ac{MC} simulation, as described in Section \ref{sec:positron_target}. Lastly, a discussion of the obtained results is reported in the conclusions (Section \ref{sec:conclusions}).

\section{Experimental Setup}
\begin{figure}[hbtp]
\centering
\includegraphics[width=8in,angle=90]{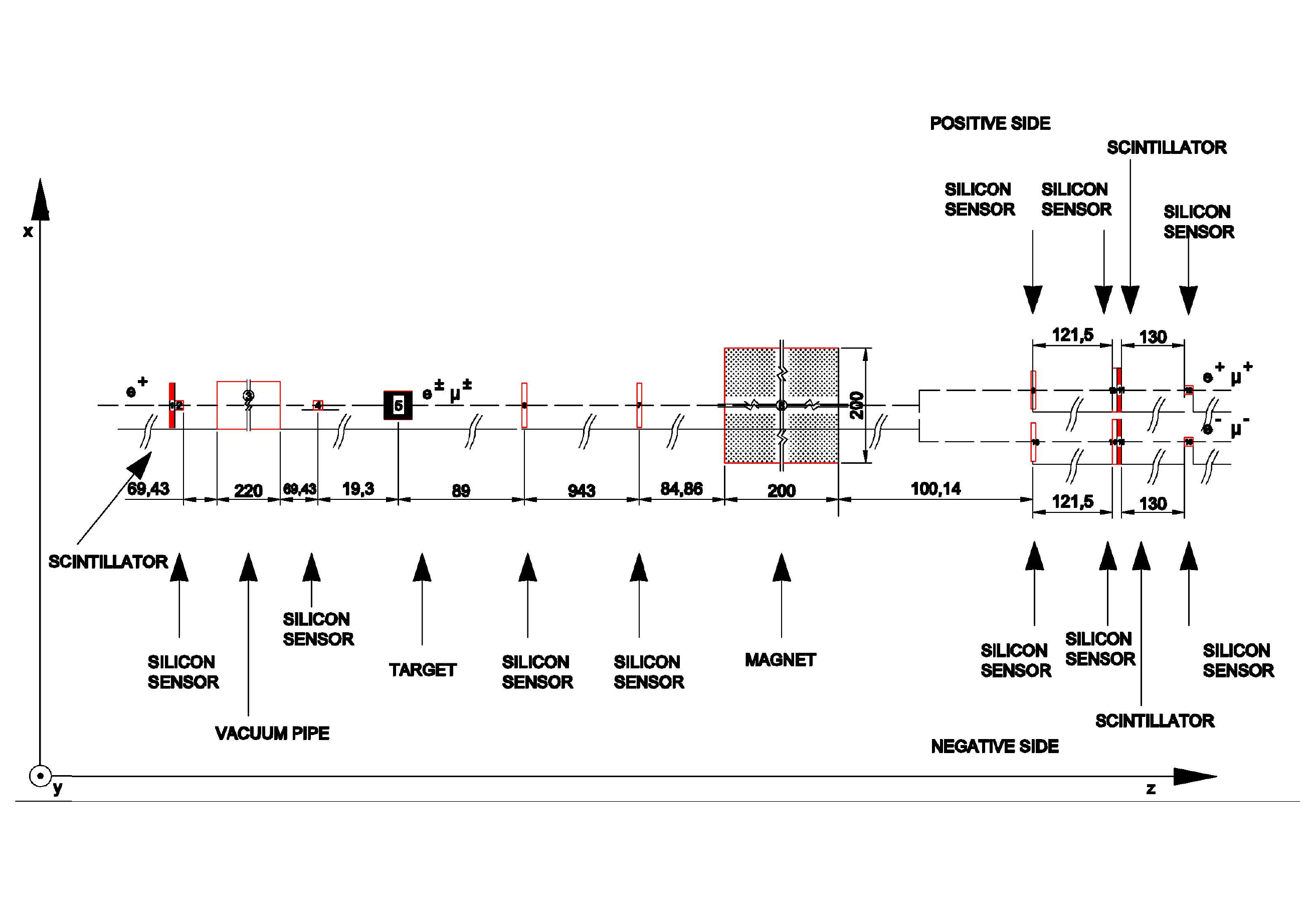}\DeclareGraphicsExtensions.
\caption{Magnetic spectrometer setup installed at the H2 beam line of \ac{CERN} to study the particles emerging from a low $Z$ target in positron annihilation processes. Lengths are expressed in $\unit{cm}$.}
\label{fig:expsetinter}
\end{figure}

\begin{figure}[hbtp]
\centering
\includegraphics[width=7in, angle=90]{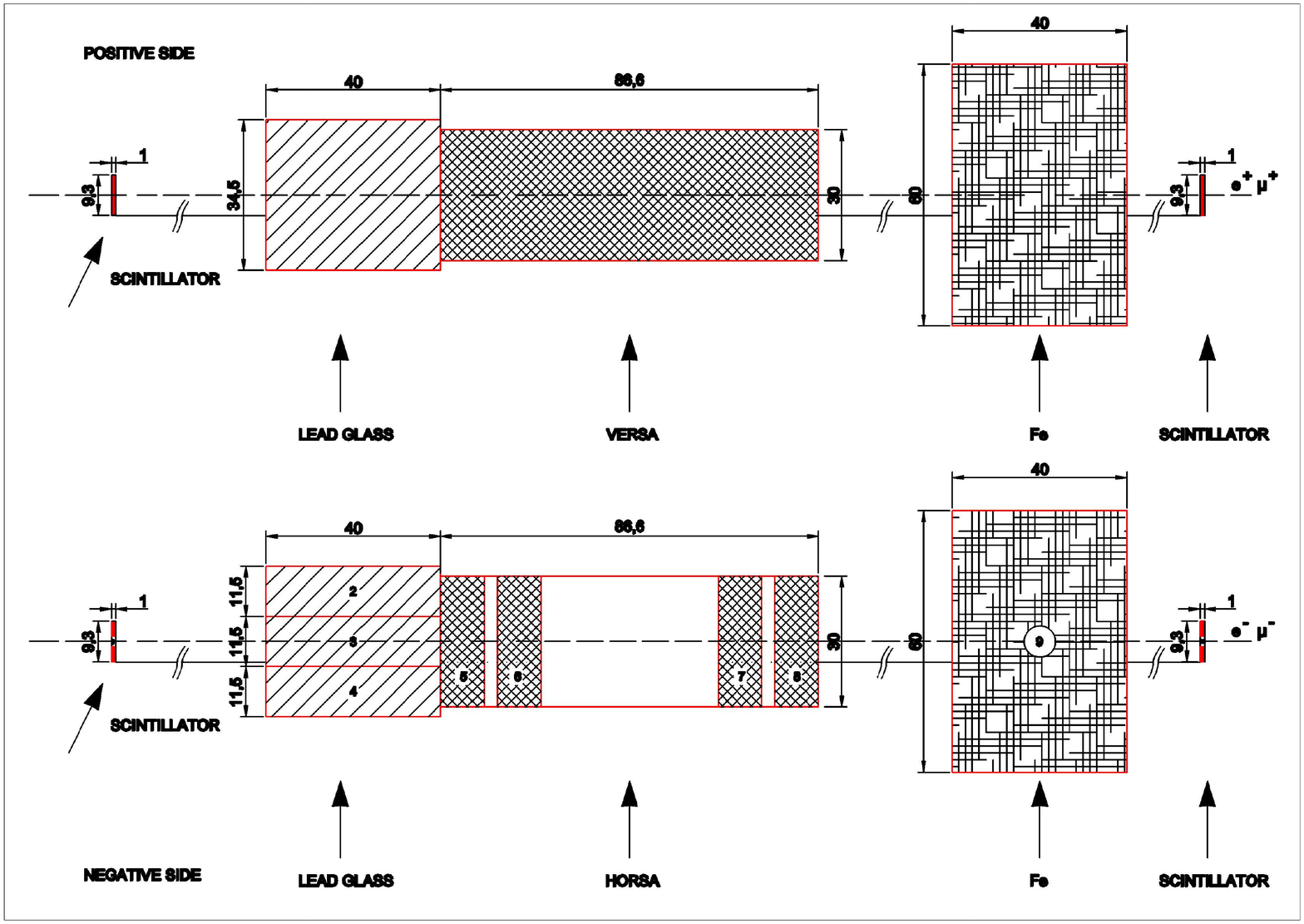}\DeclareGraphicsExtensions.
\caption{Absorbers region located downstream the magnetic spectrometer. Lengths are expressed in $\unit{cm}$.}
\label{fig:expset}
\end{figure}

The experimental setup arranged to collect data consists of {\color{black}{an approximately}} $\unit[23]{m}$ long apparatus, shown in Fig. \ref{fig:expsetinter} (vacuum pipe, target and magnetic spectrometer region) and Fig. \ref{fig:expset} (absorbers region) that was installed in the H2 beam line of the \ac{CERN} North Area during a data-taking campaign in summer 2018 to measure with high precision trajectories and momenta of the two final state muons as well as the direction of the incoming positrons. As shown in Fig. \ref{fig:expsetinter}, the coordinate system has the $z$-axis along the beam, the $x$-axis pointing upwards and $y$-axis pointing towards the reader. In both Figures \ref{fig:expsetinter} and \ref{fig:expset}, lengths are expressed in  $\unit{cm}$.

Upstream of the fixed target (Fig. \ref{fig:expsetinter}), positrons from the beam are crossing a pair of $\unit[2\times2]{cm^{2}}$ silicon sensors, constituted by two layers of orthogonal microstrips in order to measure the incoming particle direction and position. A scintillator placed upstream of the target was used for real-time event selection and data-acquisition purposes (trigger).

Another silicon sensor pair is placed downstream of the target (Fig. \ref{fig:expsetinter}) in order to measure the direction of emerging particles before these enter a dipole magnet with a  $\unit[2.01]{T}$  field along $y$, extended in a region of approximately $\pm\unit[100]{cm}$
along $z$. 

Each secondary particle (being it $e^\pm$ or $\mu^\pm$) is individually reconstructed and its momentum is determined by measuring its deflection in the $x$-$z$ plane with a two arms silicon sensor spectrometer located downstream of the magnet (Fig. \ref{fig:expsetinter}). 

After being tracked by the aforementioned spectrometer, particles are entering the absorbers region (Fig. \ref{fig:expset}) where the detectors are arranged along two arms, following the expected trajectory of positive and negative particles. Each arm is constituted by two scintillators serving as a trigger and  the massive absorbers. An iron block placed across both arms is employed to shield \ac{DT} muon chambers located further downstream (the \ac{DT} are not displayed in  Fig. \ref{fig:expset} and they have not been used in this measurement).

Each active absorber is  made of a three unit \ac{LG}
electromagnetic calorimeter,  
each unit featuring a $\unit[40]{cm}$ deep truncated pyramid shape. The \ac{LG} \ac{PMT} signals were processed by a $\unit[12]{bit}$ digitizer  \cite{Caenv1720}. The instrumented absorber is installed immediately after the \ac{LG} calorimeter. 

The absorber named \ac{HORSA}, the only device for which performances are presented in this paper,  is installed in the spectrometer arm where the negative particles (either $e^-$ or $\mu^-$) are deflected by the magnetic field. It is equipped with $\unit[1]{inch}$ thick fused silica layers where secondary charged particles are producing Cherenkov light. Three \ac{PMT}s are used to detect the light produced in each of the fused silica layers. The light is transmitted to the \ac{PMT}s  by internal reflection along the $x$ direction. 
\ac{HORSA} \ac{PMT}s were processed by a  $\unit[14]{bit}$ digitizer \cite{Caenv1730}. 
A sequence of alternating active and passive layers along $z$ are used to filter muons  against electrons. Among the first (second) pair of fused silica layers a $\unit[5]{cm}$ ($\unit[3]{cm}$) tungsten shield is inserted. Among the two fused silica layer pairs 
a passive element made of a $\unit[23]{cm}$ thick tungsten layer  is  inserted. 



Two trigger configurations were used for the shared silicon sensors and calorimeters \ac{DAQ}. A first one, named {\it single}, was based on the signal produced by the scintillator placed upstream of the target (Fig. \ref{fig:expsetinter}). A second one, named {\it muon}, was based on the coincidence of the same upstream  scintillator signal with the signals of two pairs of additional scintillators placed upstream of the \ac{LG} blocks and downstream of the \ac{DT} chambers respectively (Fig. \ref{fig:expset}), on each of the two arms of the detector (i.e. the absorber region showed in Fig. \ref{fig:expset}). 
The {\it single} trigger was used to select and record events with no bias on the final state while the {\it muon} trigger was used to enhance the content of $\mu^{+}\mu^{-}$ events in the recorded data sample.

Two kinds of \ac{DAQ} system have been used to record event information from the different subdetectors. The \ac{DAQ} system for all subdetectors, except the \ac{DT} chambers, is based on an external trigger signal given by a combination of the scintillator fast signals (the aforementioned {\it single} and {\it muon} combinations). The \ac{DT} chambers used instead a fast, $\unit[40]{MHz}$, trigger-less, \ac{DAQ} system, developed for the Run 3 data taking period of the \ac{CMS} detector \cite{CERN-LHCC-2017-014-DAQ}.  The trigger signal from the scintillators is also shared to the \ac{DT} \ac{DAQ} system for subdetector off-line synchronisation and complete event building. 

\FloatBarrier

\section{Absorbers system performance studies}\label{sec:performance}

\subsection{Data acquisition}\label{subsec:data_acquisition}
Data from events generated at the H2 beam line were recorded during the summer 2018 data-taking campaign. Electron and muon beams of different energies without the target in position were delivered to the experimental region, as well as a $\unit[45]{GeV}$ positron beam hitting the target in its position.
Muon beams with $E_0 = \unit[22]{GeV}$ energy, without any target in place, and with both magnetic field directions, have been used for the alignment of the silicon detectors and the \ac{DT} chambers. 

Muon and electron beams, both with the same energy $E_0$ and no target inserted are also used to estimate the detector's performances. This energy is approximately equal to the mean energy of the muons produced in the $e^+e^- \rightarrow \mu^+ \mu^-$ process when initiated by a $\unit[45]{GeV}$ positron beam. The magnet is operated to deflect particles with energy $E_0$ into the downstream double-arm section: particles with this energy are in fact hitting the central unit of each \ac{LG} calorimeter, depending on the sign of their charge. 
In addition to the aforementioned calibration runs, different data-taking runs were carried out with a $\unit[45]{GeV}$ positron beam delivered to different kinds of targets to study muon production close to the kinematic threshold of the $e^{+}e^{-}\rightarrow\mu^{+}\mu^{-}$ process. The results presented in this paper are based on the data collected during runs in which two Carbon targets with $\unit[2]{cm}$ and $\unit[6]{cm}$ thickness and a diameter of $\unit[4]{cm}$ were present (Sec. \ref{sec:positron_target}).
The positron beam had a pulsed shape with 4 spills per minute, each spill lasting $\unit[4.8]{s}$ with a typical intensity of $5\cdot10^{6}$ positrons per spill. The spot size was $\sim\unit[2\times2]{cm^{2}}$ with an angular spread of $\sim\unit[300]{\mu rad}$. With the chosen collimator settings the momentum spread was below $\unit[1.5]{\%}$ \cite{Brianti:19735oog} and the purity of the beam was in the range $95-99\,\%$ \cite{Charitonidis}. 

During the data-taking, information about each particle arrival time, released energy in each of the absorbers and positions detected by the silicon trackers is recorded. For the runs with the $\unit[45]{GeV}$ positron beam hitting the target, {\it muon} and {\it single} triggers were included to start the \ac{DAQ} system with a prescaling factor of about $3\cdot10^{4}$.

Events with $\mu^+\mu^-$ or $e^+e^-$ pairs in the final state were identified through the \ac{LG} and the HORSA absorber located on the negative side of the spectrometer, i.e. based only on the identification of the negative particle in the final state. Events with a $\mu^+\mu^-$ pair in the final state are mainly due to the  $e^+e^-$ annihilation process in the target. Events with a $e^+e^-$ pair in the final state are produced either in $e^+e^-$ Bhabha scattering of the impinging positron beam with an atomic electron in the target material or in pair production from a high energy photon in the target \cite{MandlAndShaw}. 

Electrons reaching the absorbers region are producing an electromagnetic shower almost entirely contained in the \ac{LG} central unit while negative muons are releasing a small amount of their energy in the detector materials hence crossing entirely the \ac{LG} and \ac{HORSA} detectors. Muons are expected to produce a signal both in the \ac{LG} central unit and in all the \ac{HORSA} active layers. Only a fraction of the energy lost by each muon is deposited as Cherenkov light and eventually detected by the \ac{PMT}s. A muon event candidate is defined based on the time coincidence of signals from the \ac{PMT}s corresponding to all the five trigger scintillators, the central \ac{LG} unit and the four \ac{HORSA} active layers,  with an anti-coincidence requirement with respect to the two external \ac{LG} units. 

Exploiting the different behaviour of muons and electrons in the apparatus, the ratio of the number of events with a muon to those with an electron, $\nicefrac{N_{\mu^{-}}}{N_{e^{-}}}$, reaching the absorber region can be computed and eventually be compared with a \ac{MC} simulation of the experimental setup.




 \subsection{Tracker - absorbers correlation} \label{subsec:tracker}
\begin{figure}[htbp]
\centering
\includegraphics[width=4in]{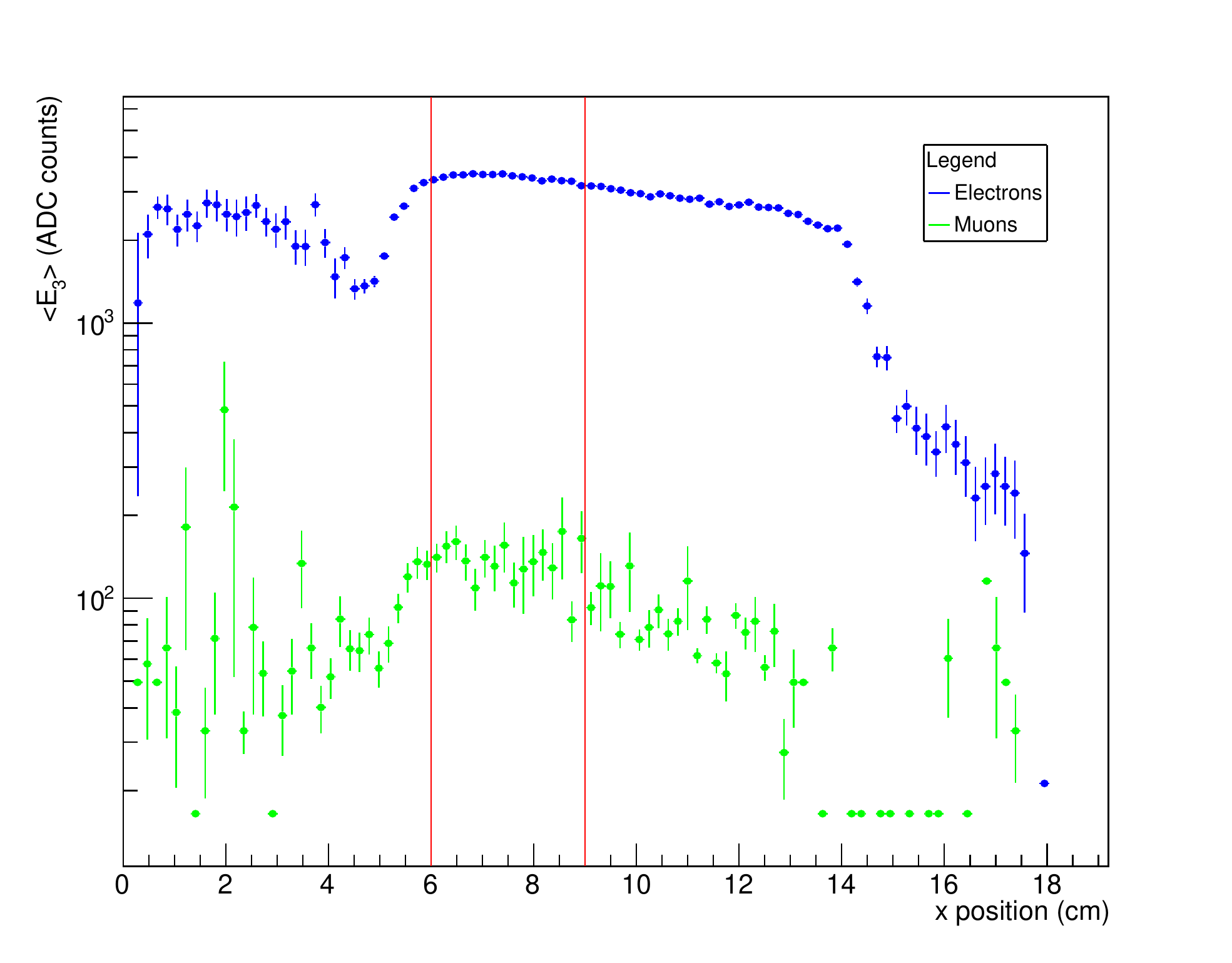}\DeclareGraphicsExtensions.
\caption{Average energy released $\left\langle E_{3}\right\rangle$  by $\unit[22]{GeV}$ electrons (blue, on the top) and muons (green, on the bottom) in the central \ac{LG} unit as a function of the local $x$ position measured by the tracker. Vertical solid lines are used to mark the region 
$\unit[6]{cm}<x<\unit[9]{cm}$, considered for further analysis.}

\label{fig:energy0-tracker}
\end{figure}

 As mentioned before, the experimental apparatus is arranged so that the magnet deflects particles with energy $E_0$ towards the central unit of the \ac{LG} (block 3 in Fig. \ref{fig:expset}). As a first step, this analysis studied the relation between the particle's $x$ position measured by the tracker (Fig. \ref{fig:expsetinter})  positioned before the first scintillator of the negative arm and the energy released $E_3$ in the central \ac{LG} unit (subscript number refers to Fig. \ref{fig:expset}). Fig. \ref{fig:energy0-tracker} shows the average electron (muon) released energy $\left\langle E_{3}\right\rangle$  in the central \ac{LG} unit in blue (green), as a function of the $x$ position measured by the tracker. The $x$-scale refers to the local coordinate system of the tracker. 
 Particles in the range 
 $\unit[6]{cm}<x<\unit[14]{cm}$  are the ones that traverse the whole central \ac{LG} unit.
 The noticeable 
 negative correlation is due to the finite energy spread of the beam. Particles crossing the tracker at lower $x$ correspond to those closer to the beam line, less deflected because of their higher momentum.
 
 
The largest released energy occurs for particles crossing the tracker with coordinates in the range 
$\unit[6]{cm}<x<\unit[9]{cm}$, both for muons and electrons. This requirement is applied to define samples of electrons or muons considered for the studies described in the following.

\subsection{Lead Glass calibration} \label{subsec:rel_ene_ele}
\begin{figure}[htbp]
\centering
\includegraphics[width=4in]{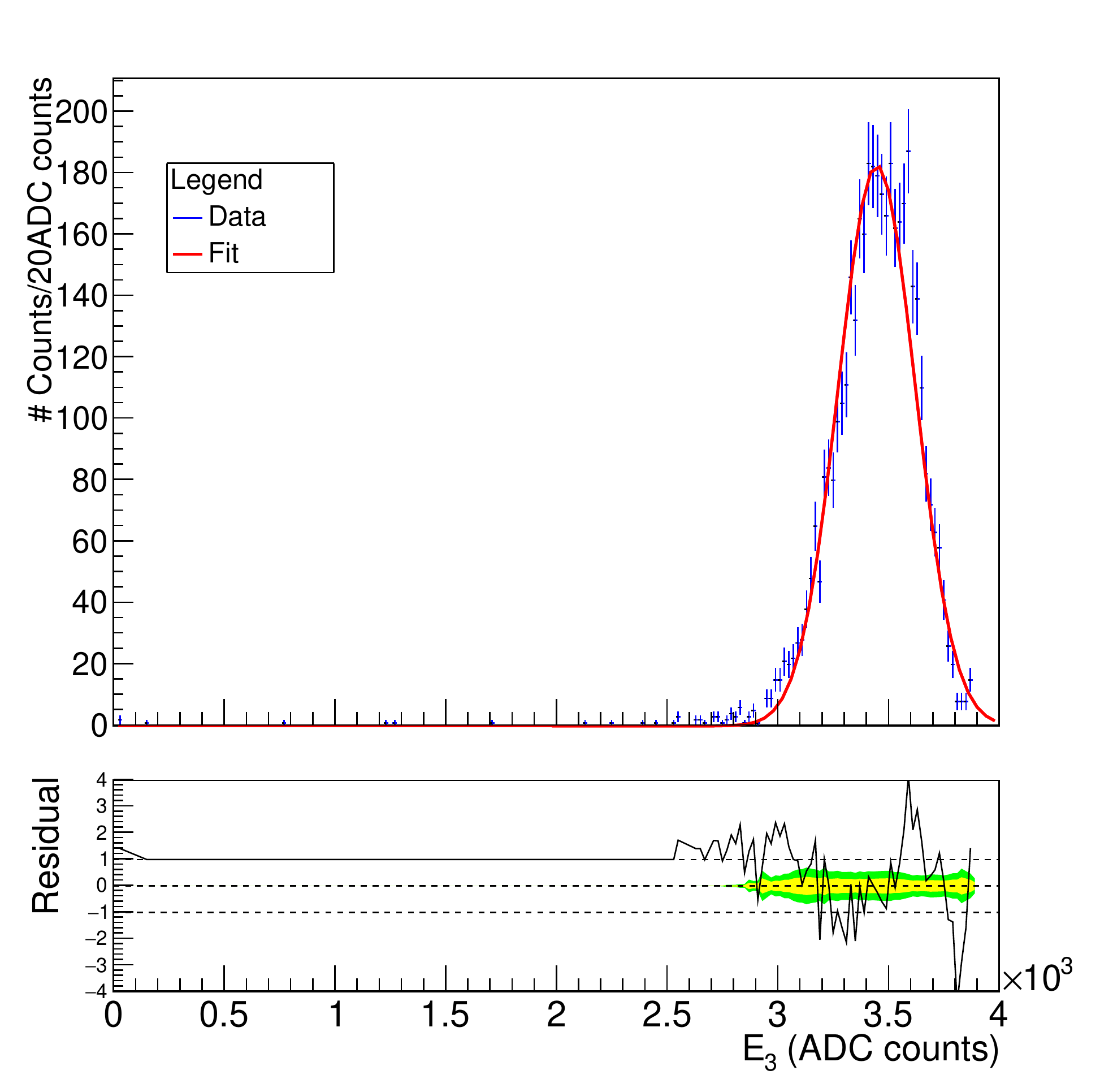}\DeclareGraphicsExtensions.
\caption{Energy released $E_3$ by electron candidates in the central \ac{LG} in a $\unit[22]{GeV}$ electron beam run with no target in place. The distribution is interpolated with a Gaussian function (solid line) of which the lower panel of the plot shows the residual. This procedure serves as calibration of the \ac{LG}.}
\label{fig:enecal}
\end{figure}

Data collected during a run with an electron beam entering the apparatus and no target in place has been considered to study the released energy in the central unit of the \ac{LG} $E_3$. This is needed to calibrate the energy response of this detector. Events with a time coincidence between signal in the scintillator (Fig. \ref{fig:expset}) and in the three \ac{LG} units  were selected. The spectrum of the energy released in the central \ac{LG} unit by the selected electron candidates is shown in Fig. \ref{fig:enecal} together with a Gaussian function interpolating the spectrum.

 The Gaussian interpolation results in a mean value of $\mu=\unit[3447]{ADC}$, the square root of the variance being $\sigma=\unit[174]{ADC}$, with a reduced chi-squared (least squares) of $\nicefrac{\chi^{2}}{\nu}=\nicefrac{175}{70}$, where $\nu=70$ is the number of the degrees of freedom. Electrons are expected to release all of their energy in the \ac{LG} volume hence it is possible to calibrate the \ac{LG} energy response according to the relation $\unit[1]{GeV} =\unit[\left(156.7\pm7.9\right)]{ADC\,counts}$, assuming a linear response. The quoted uncertainty is the statistical uncertainty, given that the systematic uncertainties on this relation are  negligible compared to the statistical one on the studies presented in the following.
 

\subsection{HOrizontal Smart Absorber efficiency} \label{subsec:HORSA_eff}
 {\color{black}The \ac{LG} calibration allows the identification of muon candidates and their discrimination from other secondary particles based on the energy they release in the central \ac{LG} unit, $E_{3}$. Muon candidates are required to have $E_{3} < \unit[1]{GeV}$ (this value  has been estimated by \ac{MC} simulations). 
 
 The  efficiency of each \ac{HORSA} active layer has been studied using a muon beam with energy $E_0$.
The efficiency of each \ac{HORSA} active layer $j=5,6,7,8$ (numbers refer to Fig. \ref{fig:expset}) is defined as:
\begin{equation}
\varepsilon_{j}=\frac{N_{4}}{N_{3}\left(j\right)}
\label{efficiency}
\end{equation}
where $N_{4}$ represents the number of events with signals in time coincidence in all four layers while ${N_{3}\left(j\right)}$ represents the number of events with signals in time coincidence in all the layers but the $j$ one.
The overall  efficiency of  the \ac{HORSA} absorber is defined as
\begin{equation}
\varepsilon_{\text{HORSA}}=\prod_{j=5}^{8}\varepsilon_{j} 
\label{Totefficiency}
\end{equation}
Results obtained from the run with a muon beam of energy $E_0$ and no target in place are given in Tab. \ref{efficiency}, together with the corresponding statistical uncertainties evaluated assuming binomial statistics.

\begin{table}[htbp]
\centering
  \caption{ Efficiencies to detect a $\unit[22]{GeV}$ muon in each of the fused silica \ac{HORSA} layers together with the overall \ac{HORSA} efficiency, as defined in Equations \ref{efficiency} and \ref{Totefficiency} respectively.}
  \label{table:HORSA_eff}
  \begin{threeparttable}
    
     \begin{tabular}{lll}
        \toprule
       $j$  & \(\unit[\varepsilon_{j}]{\left(\%\right)}\) & \(\unit[\varepsilon_{\text{HORSA}}]{\left(\%\right)} \)   \\
        \midrule
        5	&$ 97.04\pm0.56$ & \multirow{ 4}{*}{$84.2\pm1.4$}\\
        6	&$ 95.50\pm0.66$  \\
        7	&$ 93.26\pm0.98$	   \\
        8	&$ 97.44\pm0.79$	   \\
        \bottomrule
     \end{tabular}
  \end{threeparttable}
\end{table}
\subsection{HORSA electron misidentification} \label{subsec:HORSA_cont}
To evaluate the probability to mistake electrons (produced by positrons hitting a target) for muons (also produced by positrons on target),  the electron misidentification in the \ac{HORSA} absorber is evaluated using data collected in a run with an electron beam  with energy $E_0$ crossing the apparatus. 
The electron misidentification in the \ac{HORSA} absorber is given by:
\begin{equation}
e_{\text{misID}}^{-}=\frac{N_{\text{HORSA}}}{N_{\text{Tot}}}
\label{contamination}
\end{equation}
where $N_{\text{HORSA}}$ is the number of events with signals in time coincidence in the four \ac{HORSA} layers and $N_{\text{Tot}}$ is the total number of recorded events.

In the considered run of data-taking $N_{\text{HORSA}}=1$ and $N_{\text{Tot}}=21504$ so, exploiting the \ac{FC} approach for confidence interval definition, an \ac{UL} on $e_{\text{misID}}^{-}$ at $2 \cdot 10^{-4}$ has been set at $ {\,90\% \, CL}$.
Thanks to a \ac{MC} simulation of electron beam crossing the apparatus we estimate that the observed electron contamination is due to the not fully contained electromagnetic shower leaking outside of the \ac{LG} volume.

\subsection{HORSA muon detection} \label{subsec:Muon_detection}
While crossing the apparatus, muons are expected to produce a signal in the central \ac{LG} unit and in all the four \ac{HORSA} active layers.

The observable energy released by muons in the apparatus is therefore:
\begin{equation}
E_{obs}=E_{3}+E_{\text{HORSA}}.
\label{Eobs}
\end{equation}
where $E_{\text{HORSA}}=\sum_{j=5}^{8}E_{j}$ (subscript numbers refer to Fig. \ref{fig:expset}) is the released energy in the different \ac{HORSA} layers and $E_{3}$ has been already defined and is shown in Fig. \ref{fig:murel}.

\begin{figure}[htbp]
\centering
\includegraphics[width=4in]{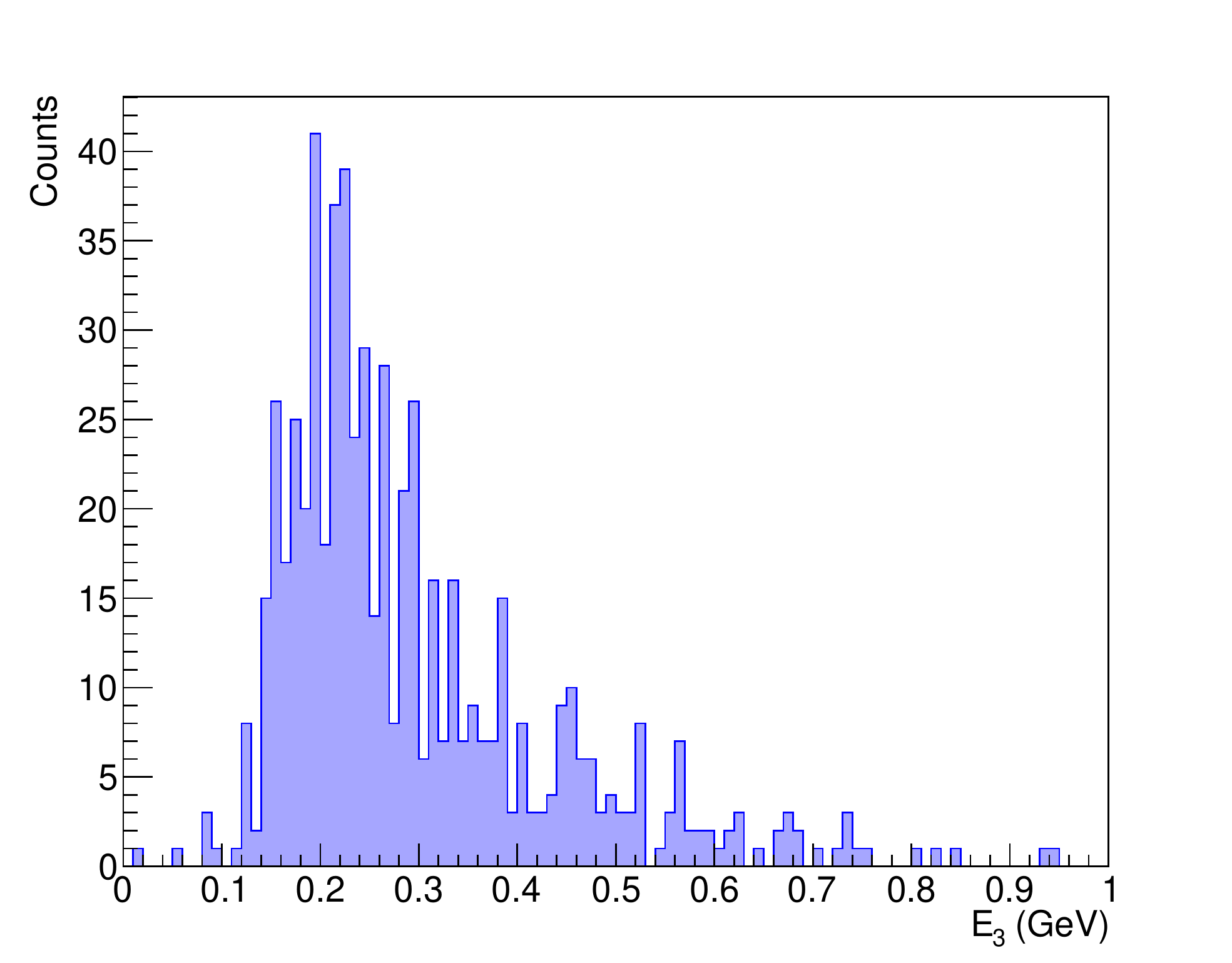}\DeclareGraphicsExtensions.
\caption{Energy released in the central \ac{LG} unit $E_{3}$, for events recorded during a \unit[22]{GeV} muon beam data-taking run. }
\label{fig:murel}
\end{figure}
 
The correlation between $E_{3}$ and $E_{HORSA}$, shown in Fig. \ref{fig:E3-EHORSA}, illustrates that the total energy released by muons with energy $E_0$ in the \ac{HORSA} detector is $E_{\text{HORSA}}< \unit[2000]{ADC}$ most of the time. This requirement can be therefore used to define a region for muon identification in \ac{HORSA} detector. This criteria will be used, together with requirements on the energy deposited in the \ac{LG} detectors, to identify muons produced in the events recorded during the \unit[45]{GeV} $e^{+}$ beam run hitting the target.

\begin{figure}[htbp]
\centering
\includegraphics[width=4in]{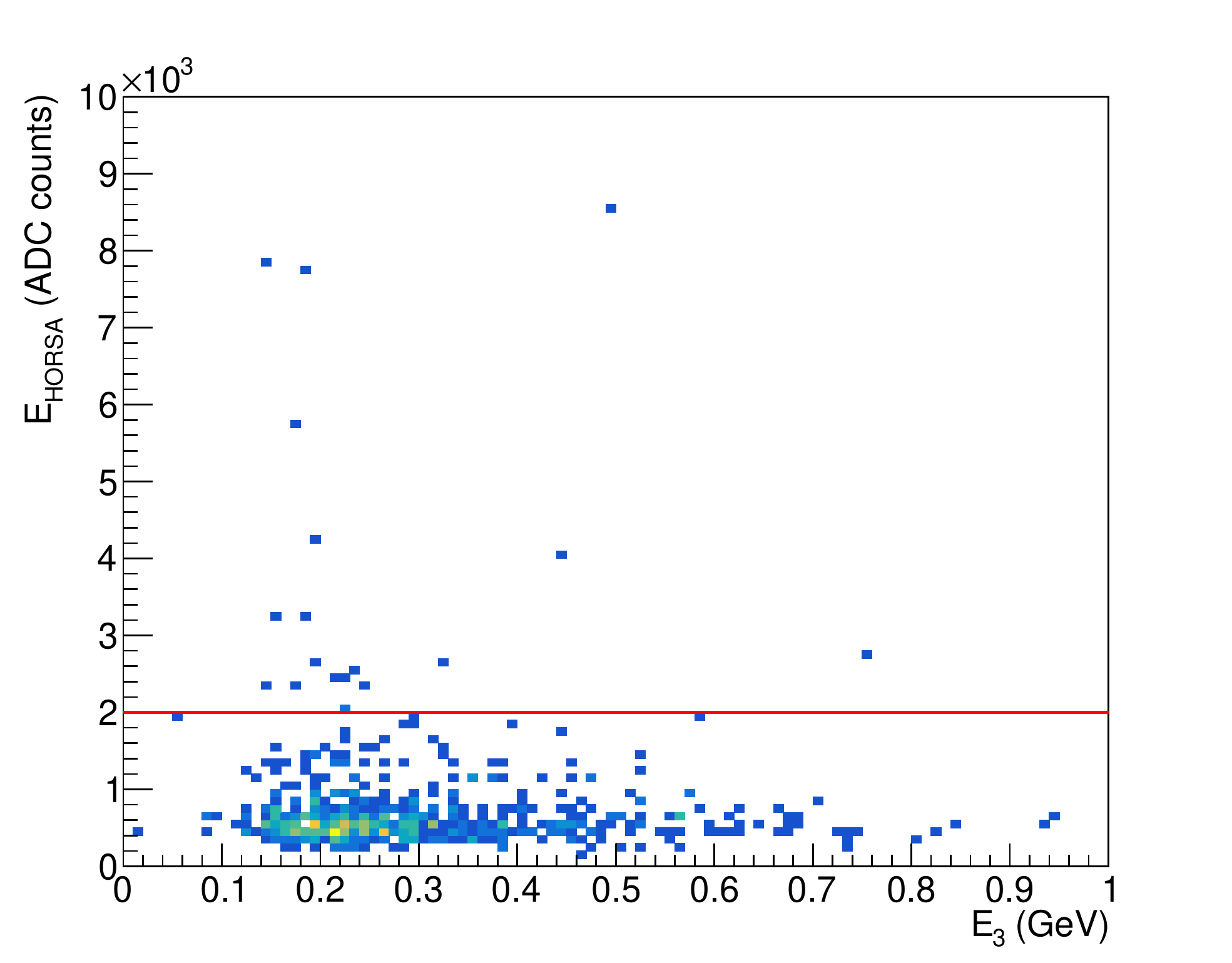}\DeclareGraphicsExtensions.
\caption{Energy released in \ac{HORSA} $E_{\text{HORSA}}$ as a function of released energy in the central \ac{LG} unit $E_{3}$ for events recorded during a \unit[22]{GeV} muon beam data-taking run. The solid line represents the requirement applied in the signal region definition.}
\label{fig:E3-EHORSA}
\end{figure}

\section{Particles detection in positron beam data with the target}\label{sec:positron_target}
The calibrated detector has been exploited to measure the $\nicefrac{N_{\mu^{-}}}{N_{e^{-}}}$ events ratio, estimating the yields of $\mu^{-}$ ($N_{\mu^{-}}$) and of $e^-$ ($N_{e^-}$), both with energy $E_{0}$. The yields are evaluated analyzing events collected during runs in which a positron beam is impinging on Carbon targets with different thicknesses, \unit[2]{cm} and \unit[6]{cm} respectively.

Considering the setup and numbering scheme shown in Fig. \ref{fig:expset}, candidate muon events have been selected requiring the time coincidence of signals in the two trigger scintillators (the first one upstream of the \ac{LG} block and the second one downstream of the muon chamber), the \ac{LG} central unit and all the \ac{HORSA} layers; anti-coincidence with the \ac{LG} external units (numbered 2 and 4 in Fig. \ref{fig:expset}) is also required. 

The contribution due to fake signal events representing a background contribution in the muon identification region, defined requiring $E_{\text{HORSA}} < \unit[2000]{ADC}$ is evaluated using the sideband-substraction
method. An orthogonal sideband region is defined by requiring $E_{\text{HORSA}} \geq \unit[2000]{ADC}$.
Assuming a uniform background distribution across the considered $E_{\text{HORSA}}$ range, the signal yields after the sideband-substraction are given by:
\begin{equation}
N_{\mu_{det}^{-}}=N_{\mu^{-}}-\alpha N_{side}
\label{N_mudet}
\end{equation}
where $N_{\mu^{-}}$  and $N_{side}$ are the number of candidate muon events in the signal region and the number of background events in the sideband region, respectively. 
The parameter $\alpha$ takes into account the different width of the signal and sideband region; this has been fixed to 
\begin{equation}
\alpha=\frac{\unit[2000]{ADC}}{E_{max}-\unit[2000]{ADC}}
\label{alpha_param}
\end{equation}

 where $E_{max}$ is the upper edge of the sideband region, being approximately $\unit[30000]{ADC}$ in the two considered runs.

Considering the setup and numbering scheme shown in Fig. \ref{fig:expset}, an electron from the $e^+e^-$ final state is selected requiring the time coincidence of signals in the two trigger scintillators (one for each arm of the detector) positioned upstream of the \ac{LG} blocks and the central unit of the \ac{LG} in the negative arm. Time anti-coincidence with the signal from the two trigger scintillators (one for each arm of the detector) positioned downstream of the muon chambers is also required.

Defining as $N_{e_{det}^{-}}$ and $\varepsilon_{e^{-}}$ the number of detected electrons and the efficiency to detect them, respectively, and taking into account the evaluated muon detection efficiency $\epsilon_{HORSA}$, the ratio between the produced muons and electrons is:
\begin{equation}
\frac{N_{\mu^{-}}}{N_{e^{-}}}=\frac{N_{\mu_{det}^{-}}}{\varepsilon_{\text{HORSA}}}\cdot\frac{\varepsilon_{e^{-}}}{N_{e_{det}^{-}}}
\label{Nmuprod/epsilon/Ne/epsilon}
\end{equation}

The efficiency to detect an electron has been estimated being close to unity through a \ac{MC} simulation. Therefore, the overall efficiency ratio can be assumed to be $\varepsilon \sim \varepsilon_{\text{HORSA}}$ and $N_{e_{det}^{-}} \sim N_{e^{-}}$. Under these assumptions Eq. \ref{Nmuprod/epsilon/Ne/epsilon} can be reduced to:

\begin{equation}
\frac{N_{\mu^{-}}}{N_{e^{-}}}=\frac{1}{\varepsilon_{\text{HORSA}}}\cdot\frac{N_{\mu_{det}^{-}}}{N_{e^{-}}}~.
\label{Nmuprod/Ne}
\end{equation}

The measured $\nicefrac{N_{\mu^{-}}}{N_{e^{-}}}$ ratios for runs corresponding to Carbon targets of different thicknesses exposed to the positron beam are summarized in Tab. \ref{tab:tab_Nmuproduced}. The uncertainties on the ratios have been evaluated propagating statistical uncertainties on $N_{\mu_{det}^{-}}$ and $N_{e^{-}}$, assuming a Gaussian uncertainty when counts were $>$ 30 and a Poisson uncertainty otherwise.
Several source of systematic uncertainties have been taken into account. The majority of the systematic effects are expected to affect the electron and muon detection in a similar way, hence will cancel in the ratio at first order. The limited amount of recorded muon events prevented from further investigations on additional systematic effects.

\begin{table}[htbp]
\centering
\caption{Observed electron and muon yields together with the corresponding ratio $\nicefrac{N_{\mu^{-}}}{N_{e^{-}}}$ for two runs corresponding to different Carbon target thicknesses.}
\label{tab:tab_Nmuproduced}
  \begin{threeparttable}
     \begin{tabular}{lllll}
        \toprule
       Thickness \(\left(\unit{cm}\right)\) & \(N_{e^{-}}\)  & \(N_{\mu_{det}^{-}}\) & \(\nicefrac{N_{\mu^{-}}}{N_{e^{-}}}\) \\
        \midrule
          \(2\)&	$85.0 \pm 9.2$ & $37.0\pm6.6$  &$\left(5.1\pm1.1 \right)\cdot10^{-1}$ 	   \\
        \(6\)&	$78.0 \pm 8.8$ & $5.0^{+4.1}_ {- 3.0}$  &$\left(7.6^{+6.4}_ {- 4.6} \right)\cdot10^{-2}$ 	   \\
       
        \bottomrule
     \end{tabular}
  \end{threeparttable}
\end{table}
\subsection{Data to Monte Carlo simulations comparisons}
The $e^{+}e^{-}\rightarrow\mu^{+}\mu^{-}$ cross section is known to be significantly smaller than the $e^{+}e^{-}\rightarrow e^{+}e^{-}$ one, with a ratio of approximately $\frac{1}{8}$ for CM energies close to the kinematic threshold for muon pair production. This value has been obtained from the BabaYaga \ac{MC} generator \cite{Babayaga} using positrons and free electrons at rest in the initial state. 

In order to validate the absorber system particle detection ability leading to the results shown in Tab. \ref{tab:tab_Nmuproduced}, a \textsc{Geant4} \cite{Geant4.site, Agostinelli:2003oog} based \ac{MC} simulation of particle electromagnetic interactions with matter has been developed, to compare the $\nicefrac{N_{\mu^{-}}}{N_{e^{-}}}$ events ratio obtained in data with theoretical predictions.

 Positrons beams along the $z$-axis, made of $10^{10}$ $\unit[45]{GeV}$ primary $e^{+}$ with characteristics equivalent to the experimental ones and  hitting Carbon targets (both with \unit[2]{cm} and \unit[6]{cm} thickness) have been simulated in order to evaluate the expected number $N_{e_{MC}^{-}}$ of produced $e^{-}$ and the $N_{\mu_{MC}^{-}}$ of produced $\mu^{-}$.
 
 
 The electron and muon yields predicted by the simulations, as well as their ratio are reported in Tab. \ref{tab:tab_MC_ratio} for the two target thicknesses considered. 
\begin{table}[htbp]
\centering
 \caption{Predicted electron and muon yields together with the corresponding ratio $\nicefrac{N_{\mu^{-}}}{N_{e^{-}}}$ for two simulated runs corresponding to different Carbon target thicknesses.} 
 \label{tab:tab_MC_ratio} 
  \begin{threeparttable}
     \begin{tabular}{llll}
        \toprule
        Thickness \(\left(\unit{cm}\right)\) & \(N_{e_{MC}^{-}}\)  & \(N_{\mu_{MC}^{-}}\)  & \(\nicefrac{N_{\mu_{MC}^{-}}}{N_{e_{MC}^{-}}}\)  \\
        \midrule
	\(2\)     &$2400 \pm 49$    &$1481 \pm 38$   &$\left(6.2\pm1.3\right)\cdot10^{-1}$   \\
\(6\)    &$\left(521\pm2.3\right)\cdot10^{2}$    &$2870 \pm 54$   &$\left(5.51\pm0.26\right)\cdot10^{-2}$   \\

        \bottomrule
     \end{tabular}
  \end{threeparttable}
\end{table}
The experimental results reported in Tab. \ref{tab:tab_Nmuproduced} are well in agreement, within uncertainties, with the corresponding predictions from \ac{MC} simulations, reported in Tab. \ref{tab:tab_MC_ratio}.

\section{Conclusions}
\label{sec:conclusions}
The implementation of the \ac{LEMMA} muon collider concept needs an intense positron beam to produce muons at a significant rate. The knowledge of the muon production cross section close to its kinematic threshold is extremely important for a muon collider design based on this concept. 

In this paper we reported the results of a muon identification technique based on segmented massive absorbers that can be used in experiments to measure the  properties of the muons produced by the $e^{+}e^{-}$ annihilation at kinematic threshold of muon pair production. 
Detected events allowed testing the performance of the considered absorbers system. A comparison with a \ac{MC} simulation resulted in a good agreement between experimental observations and predictions. This apparatus and the described particle identification method can be further deployed for future studies of the $e^{+}e^{-}\rightarrow\mu^{+}\mu^{-}$ process at its kinematic threshold energy. 


In order to increase statistics and reduce backgrounds, i.e. to provide a meticulous measurement of $e^{+}e^{-}\rightarrow\mu^{+}\mu^{-}$ cross section, more accurate tracking devices, alignment infrastructures and a more efficient trigger and readout systems will be needed. An upgrade of the \ac{LEMMA} experimental layout is foreseen in order to accumulate more data at the muon pair production threshold energy. 

With an adequately low energy spread and with a larger beam rate this apparatus might also be employed to search for bound states as the true muonium \cite{Brodsky:2009gx}.

\section*{Acknowledgements}
\addcontentsline{toc}{section}{Acknowledgements}

We would like to warmly thank the \ac{SPS} staff and the \ac{LSM} group, in particular Henrik Wilkens and Nikolaos Charitonidis, for their support during installation and data taking. This work has been supported by the ERC CoG GA 615089 CRYSBEAM.


\acrodef{ADC}[ADC]{Analogue to Digital Converter}
\acrodef{bSM}[bSM]{beyond the Standard Model}
\acrodef{CERN}[CERN]{European Organization for Nuclear Research}
\acrodef{CL}[CL]{Confidence Level}
\acrodef{CM}[CM]{center-of-mass}
\acrodef{CMS}[CMS]{Compact Muon Solenoid}
\acrodef{DAQ}[DAQ]{Data Acquisition}
\acrodef{DT}[DT]{Drift Tube}
\acrodef{FC}[FC]{Feldman-Cousins}
\acrodef{FCC}[FCC]{Future Circular Collider}
\acrodef{HEP}[HEP]{High Energy Physics}
\acrodef{HCAL}[HCAL]{Hadron CALorimeter}
\acrodef{HORSA}[HORSA]{HOrizontal Smart Absorber}
\acrodef{INFN}[INFN]{National Institute for Nuclear Physics}
\acrodef{LEMMA}[LEMMA]{Low EMittance Muon Accelerator}
\acrodef{LG}[LG]{Lead Glass}
\acrodef{LHC}[LHC]{Large Hadron Collider}
\acrodef{LSM}[LSM]{Large Scale Metrology}
\acrodef{MAP}[MAP]{Muon Accelerator Program}
\acrodef{MC}[MC]{Monte Carlo}
\acrodef{NIM}[NIM]{Nuclear Instruments and Methods}
\acrodef{PMT}[PMT]{photomultiplier}
\acrodef{QED}[QED]{Quantum ElectroDynamics}
\acrodef{SPS}[SPS]{Super Proton Synchrotron }
\acrodef{SM}[SM]{Standard Model}
\acrodef{UL}[UL]{Upper Limit}



\bibliography{mybiblio}

\end{document}